\documentstyle[aps,epsf,multicol]{revtex}
\begin{document}
\draft

\title{A model for the production of regular fluorescent light from 
coherently driven
atoms}
\author{K. Jacobs}
\address{Optics Section, The Blackett Laboratory, Imperial College, 
London SW7 2BZ, England}
\maketitle

\begin{abstract}
It has been shown in recent years that incoherent pumping 
through multiple atomic
levels provides a mechanism for the production of highly anti-bunched 
light, and that
as the number of incoherent steps is increased the light becomes 
increasingly
regular. We show that in a resonance fluorescence situation, a 
multi-level
atom may be multiply coherently driven so that the fluorescent 
light is highly
anti-bunched. We show that as the number of coherently driven 
levels is increased, 
the spontaneous emissions may be made increasingly more 
regular. We present a
systematic method for designing the level structure and driving 
required to produce
highly anti-bunched light in this manner for an arbitrary even 
number of levels.
\end{abstract}

\pacs{42.50.Ar, 42.50.Lc, 42.50.Dv}

\begin{multicols}{2}

\section{Introduction} \label{sec1}
It is well known that resonance fluorescence from a two level 
atom is 
anti-bunched~\cite{CandW,Carm}. That is, consecutive fluorescent 
photons are less
likely to arrive very close together (ie. to be bunched) than they 
are in a classical
light beam of the same average intensity. The distribution of 
waiting-times between
photo-detections for resonance fluorescence is therefore peaked 
further away from
zero than for classical light. The reason for this effect is easily 
understood. When
the atom spontaneously emits to produce a fluorescent photon it 
places itself in the
ground state. It cannot emit again until the laser field driving the 
atom has increased
the occupation probability of the excited state. The evolution of 
the excited state
occupation probability amplitude for a driven two level atom 
(excluding spontaneous
emission) is given in Fig.(\ref{fig1}). It is the nature of this 
evolution (along with the
spontaneous emission rate of the excited state) which determines 
the form of the
waiting-time distribution.

We note that in recent years it has been shown that in laser 
models with more than
two levels,  it is possible to achieve sub-Poissonian statistics in the
output~\cite{subpmisc}. In general sub-Poissonian statistics 
implies
anti-bunching~\cite{WandMetc}. Ritsch {\em et al.}~\cite{RGZW} 
have shown that
highly sub-Poissonian light may be produced if the atomic state is 
recycled from the
lower to the upper lasing level via a large number of incoherent 
pumping steps
through intermediate levels. As the number of pumping steps 
tends to infinity the
intensity fluctuations of the output light tends to zero. In this 
paper we will consider
a single atom resonance fluorescence situation rather than a laser 
model. In this case
the driving is completely coherent, and the anti-bunched light is 
produced by a single
incoherent transition.

It has been shown previously that a coherently driven three level 
Raman laser gives a
sub-Poissonian output~\cite{CohRam}. Also, Schernthanner and
Ritsch~\cite{anotherR} have shown that detuning of the pump and 
laser light from the
atomic transitions in a Raman (3-level) laser can lead to a value of 
the Mandel Q
parameter close to $-1$, and hence a highly sub-Poissonian 
output. Ralph and
Savage~\cite{RandS} have shown that a coherently driven 4-level 
laser will produce
amplitude squeezed light. In this paper we treat atoms with an 
arbitrary, but even,
number of levels.

We might call the limit of extreme anti-bunching {\em regular} 
light. That is, light in
which the time between consecutive  photon-detections is 
constant, or at least, the
fluctuations in the  temporal separation of consecutive photons is 
small compared to 
the average value of this separation. Clearly, if we can arrange the 
atomic evolution
so that the occupation probability of an unstable  state makes a 
sharp transition
from zero to some nonzero value a  well defined time after a 
spontaneous emission,
then if we have a high spontaneous emission rate the result will 
be essentially
regular fluorescence.

In the following section we show how we may design a coherently driven 
$n$ level 
atomic system
(where $n$ is even) to produce an evolution in which the anti-bunching 
is increased
over that for a two level system. The evolution may  be made 
increasingly closer to a
given desired evolution (for  example that to produce regular 
light) as the number of
levels is  increased. In Section~\ref{sec3} we give an example of 
this procedure for
the case of a four-level system. In  Section~\ref{sec4} we calculate 
the waiting-time
distributions  achieved for various $n$. We conclude in 
Section~\ref{sec5}.

\section{Tailoring the Atomic Evolution} \label{sec2}
Excluding spontaneous emission, the equation of motion for the 
occupation probability amplitudes for a coherently driven $n$ level atomic 
system may be written~\cite{WandM,CCT}
\begin{equation}
  \frac{d}{dt}\left( \begin{array}{c} a_1 \\ \vdots \\ a_n
\end{array} \right) = -iH \left( \begin{array}{c} a_1 \\ \vdots \\ 
a_n \end{array} \right) ,
\end{equation}
where $a_j$ is the probability amplitude for state $j$, and $H$ is 
an $n\times n$ Hermitian matrix, whose elements, $h_{ij}$, we will take to be real for simplicity. 
We are also using units
\setlength{\parindent}{0em} 
\begin{figure}
\begin{center}
\leavevmode
\epsfxsize=6cm 
\epsfbox{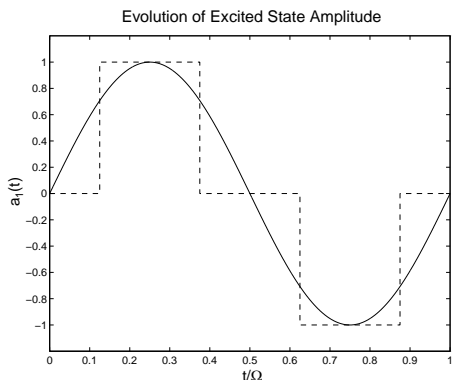}
\caption{\narrowtext Solid line: The evolution of the excited state probability 
amplitude of a driven two level atom with Rabi frequency 
$\Omega$. 
Dashed line: An imaginary evolution that would produce light, the 
regularity of
which would be limited only by the spontaneous decay rate of the 
excited state.}
\label{fig1}
\end{center}
\end{figure}
such that $\hbar=1$. The matrix $H$, which is the 
Hamiltonian, is determined both by the laser fields driving the atomic levels, and the dipole matrix elements between the atomic levels. The 
magnitude of the off-diagonal element $h_{ij}$ is given by the product of the 
strength of the laser field
 coupling the atomic levels $i$ and $j$ with the magnitude of the 
dipole matrix element connecting those levels. The phase of the off-diagonal element $h_{ij}$ is determined 
by the the phase of the respective coupling laser field and the sign of the respective dipole element. The diagonal elements are determined 
by the detuning of the laser fields from the various atomic 
transitions being driven~\cite{Agar,Loui}. 
The general evolution (up until any spontaneous decay) is given by
\begin{equation}
 \left( \begin{array}{c} a_1(t) \\ \vdots \\ a_n(t) \end{array} 
\right) = \sum_{m=1}^n c_m e^{-i\lambda_m t} {\bf v}_m  \; , \;\; 
H{\bf v}_m = \lambda_m{\bf v}_m ,
\end{equation}
where the coefficients $c_m$ are determined by the initial conditions 
chosen for the $a_j$. The ${\bf v}_m$ are the eigenvectors of 
$H$. We will write their components as ${\bf v}_m=(v_{1m}, 
\ldots ,v_{nm})$, and assume them to be normalised so that ${\bf 
v}_m\cdot {\bf v}_l = \delta_{ml}$.\setlength{\parindent}{1em}

We now impose certain conditions on the laser interaction Hamiltonian $H$. 
We will show below how interaction matrices satisfying these conditions may 
be constructed. Once the interaction matrix has been constructed, this tells
us the configuration of laser fields that is required to drive the atomic 
levels to produce the desired evolution. First of all let us assume that 
$n$ is even, and that for every eigenvalue of $H$ there is another 
eigenvalue which has the same magnitude but opposite sign, so 
that the eigenvalues come in pairs which sum to zero. Let us take 
the initial condition to be $a_n=1$, and all the other amplitudes 
zero, so that the atom starts initially in the state $n$. Finally, let 
us 
assume that the $n$ values $c_mv_{1m}$ also come in opposite 
pairs, such that $c_mv_{1m}-c_l v_{1l}= 0$ iff $\lambda_m-
\lambda_l = 0$. The evolution of $a_1$ may now be written
\begin{equation}
  a_1(t) = \sum_{m=1}^{n/2} 2i c_mv_{1m} \sin(\lambda_m t), 
\label{sinser}
\end{equation}
which is clearly a sine series for the evolution of $a_1$. 

Before we show how to construct a matrix $H$ with the above 
characteristics, let us turn to the question of regular photon 
emission. Let us assume that all the levels in our system have 
much lower spontaneous decay rates than level 1, and that this 
level decays via spontaneous emission to level $n$. For regular 
photon emission we therefore require that the evolution of the 
amplitude of level 1, given by the (finite) sine series above, have 
a 
sharp transition from zero to some non-zero value. An ideal 
evolution would be given, for example, by the dotted `square' 
curve in Fig.(\ref{fig1}). To approximate the ideal curve to a given 
level of accuracy we may choose the sine series in 
Eq.(\ref{sinser}) to be the first $n$ terms in the Fourier sine series 
for the desired `square' evolution. The Fourier series in question is
\begin{equation}
  f(t) = \sum_{m=1}^\infty \frac{2\sqrt{2}\mbox{Im}[(1+i)i^m]} 
{(2m-1)\pi} \sin{((2m-1)2\pi\Omega t)} , \label{fser}
\end{equation}
where $1/\Omega$ is the period of f(t). Clearly the evolution will 
more closely follow the desired evolution as $n$ is increased.

We should note here that we only require the evolution to be 
equal to the  Fourier
series up to a constant factor. This is because the overall scaling of 
the evolution is
not important as far as the resulting waiting time distributions are 
concerned; we
can cancel the effect of scaling the evolution curve by scaling the 
spontaneous
emission rate of the unstable level with respect to $\Omega$. This 
will be obvious
when we write down the expression for the waiting time 
distribution in
Section~\ref{sec4}.

To construct the required matrix $H$ we first write it in the form
\begin{equation}
   H= TDT^t
\end{equation}
where $D$ is a diagonal matrix containing the eigenvalues of $H$, 
and T is a matrix whose columns are the normalised eigenvectors 
of $H$. Constructing $H$ involves choosing the elements of $D$ 
and $T$. We are clearly free to choose the eigenvalues of $H$, by 
choosing the diagonal elements of $D$. The second condition 
imposed above
concerns the $n$ values $c_mv_{1m}$. Now the $c_m$ depend 
upon the initial conditions. Recall that we choose the initial 
conditions to be $a_1=\ldots=a_{n-1}=0,a_n=1$. With this choice it 
is easily shown that $c_m=v_{nm}$. That is, the coefficient $c_m$ 
is given by the last element of vector ${\bf v}_m$; the 
condition on the coefficients is a condition on the elements of the 
eigenvectors of $H$. In particular it is a condition on the first and 
last elements of the eigenvectors.

It is pertinent to note now that if the columns of a square matrix 
are orthonormal vectors, it follows that the rows are also 
orthonormal. To 
construct $H$ we therefore require both to find a set of $n$ 
orthonormal rows for $T$, and satisfy the condition on the first 
and last elements of the vectors ${\bf v}_m$. The first elements of 
the ${\bf v}_m$ are the $n$ elements of row one, and the last 
elements of the ${\bf v}_m$ are the $n$ elements of row $n$. The 
condition 
we must satisfy is that the $n$ values 
$v_{nm}v_{1m}$ must sum to zero in pairs, which clearly implies
\begin{equation}
  \sum_{m=1}^n v_{nm}v_{1m} = 0.
\end{equation}
This simply states that the dot product of row one with row $n$ 
should vanish. The condition 
on the elements of the ${\bf v}_m$ is therefore consistent with 
the orthogonality of the rows.

The $v_{1m}$ and $v_{nm}$ may now be chosen so that the 
coefficients in the sine series for the evolution of $a_1$ match 
(up to an overall constant factor) the desired first $n$ Fourier 
coefficients. Choosing the other $n-2$ rows of $T$ is then a 
straight forward procedure in linear algebra. (For example the 
Gramm-Schmidt orthogonalisation procedure may be 
used~\cite{GS}).

Note that there is clearly a degree of freedom in constructing the 
matrix $H$. If we take the matrix $T$ to be real, then there are 
$n^2$ undetermined coefficients. Orthonormality of the rows 
imposes $n(n+1)/2$ conditions. Obtaining the desired Fourier 
coefficients imposes $(n-1)$ conditions (one less than $n$ because 
the overall scaling of the coefficients is unimportant). This leaves 
$(n/2-1)(n-1)$ elements undetermined. These may therefore in 
general be chosen so as to simplify the form of $H$ in order to 
simplify the corresponding physical system. 

\section{An Example for a Four-Level System} \label{sec3}
We now use the procedure described in the previous section. We 
construct a Hamiltonian for a four-level system that will produce 
evolution corresponding to the first two terms in the sine series 
given by Eq.(\ref{fser}).

\begin{figure}
\begin{center}
\leavevmode
\epsfxsize=6cm 
\epsfbox{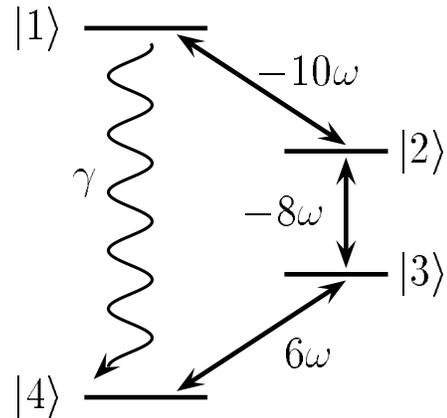}
\caption{\narrowtext A diagrammatic representation of the driven four-level 
atomic 
system designed to produce light with anti-bunching increased 
over a two level system. The elements of the Hamiltonian coupling 
the various levels
are given as multiples of $\omega=2\pi\Omega 20^{-1/2} $ . }
\label{fig2}
\end{center}
\end{figure}

First we choose the eigenvalues $-6\pi\Omega$, $-
2\pi\Omega$, $2\pi\Omega$, $6\pi\Omega$, which give us 
respectively the 
diagonal elements of $D$. To satisfy the condition on the elements 
of $T$ we first choose the top row of $T$ to have elements all 
identical. As a result the bottom row of $T$ must be chosen to 
have elements which vanish in pairs, the absolute value of each 
pair being proportional to the absolute value of a Fourier 
expansion 
coefficient. Thus 
we choose as the bottom row $(1,-3,3,1)/\sqrt{20}$. The other 
two rows are  free to be any two further mutually orthonormal 
vectors. We decide to choose them such that three off diagonal 
elements of $H$ are zero. The resulting $T$ matrix is
\begin{equation}
  T = \left( \begin{array}{cccc} 
         \mbox{\scriptsize $1/2$} &          \mbox{\scriptsize $1/2$} &         \mbox{\scriptsize $1/2$} & \mbox{\scriptsize $1/2$} \\
\mbox{\scriptsize $-3/\sqrt{20}$} & \mbox{\scriptsize $-1/\sqrt{20}$} & \mbox{\scriptsize $1/\sqrt{20}$} & \mbox{\scriptsize $3/\sqrt{20}$} \\
         \mbox{\scriptsize $1/2$} &         \mbox{\scriptsize $-1/2$} &        \mbox{\scriptsize $-1/2$} & \mbox{\scriptsize $1/2$} \\
 \mbox{\scriptsize $1/\sqrt{20}$} & \mbox{\scriptsize $-3/\sqrt{20}$} & \mbox{\scriptsize $3/\sqrt{20}$} & \mbox{\scriptsize $-1/\sqrt{20}$} 
\end{array} \right) .
\end{equation}
The resulting Hamiltonian is given by
\begin{equation}
  H = T D T^t  = \frac{2\pi}{\sqrt{20}}\Omega\left( 
\begin{array}{cccc} 
  \mbox{\scriptsize $0$} & \mbox{\scriptsize $-10$} &  \mbox{\scriptsize $0$} & \mbox{\scriptsize $0$} \\
\mbox{\scriptsize $-10$} &   \mbox{\scriptsize $0$} & \mbox{\scriptsize $-8$} & \mbox{\scriptsize $0$} \\
  \mbox{\scriptsize $0$} &  \mbox{\scriptsize $-8$} &  \mbox{\scriptsize $0$} & \mbox{\scriptsize $6$} \\
  \mbox{\scriptsize $0$} &   \mbox{\scriptsize $0$} &  \mbox{\scriptsize $6$} & \mbox{\scriptsize $0$}
\end{array} \right) .
\end{equation}
Note that all the diagonal terms are zero. While the off-diagonal 
elements describe
the coupling of the atomic energy levels via classical driving 
fields, the diagonal
elements are determined by the detuning of the driving fields 
from the atomic
transitions which they drive. The diagonal
elements of the Hamiltonian are zero when all of the detunings are zero, 
so that the Hamiltonian we have 
constructed describes the situation where all the lasers are tuned to 
the transitions they drive. The value of $\Omega$ determines the 
strength of the laser fields, and may clearly be thought of as a 
generalised Rabi frequency. In Fig.(\ref{fig2}) we give a diagrammatic 
representation of the four-level atomic system showing the coupling 
between the levels.

\section{Waiting-Time Distributions} \label{sec4}
\begin{figure}
\begin{center}
\leavevmode 
\epsfxsize=6cm 
\epsfbox{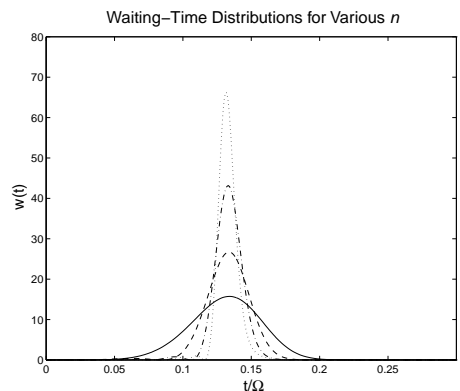}
\caption{\narrowtext Waiting-time distributions that would be generated by 
systems which produce an evolution matching the first $n$ terms 
in the Fourier series given by Eq.(\protect\ref{fser}), for various 
values of 
$n$. Solid line: $n=2$. Dashed line: $n=4$. Dash-dot line: $n=8$. 
Dotted line: $n=16$. The value of the spontaneous emission rate
is $\gamma=100\Omega$. }
\label{fig3}
\end{center}
\end{figure}

We now derive the waiting time distribution from the 
evolution of the unstable state. Appealing to the Monte-Carlo 
Wave Function (or `quantum trajectory') method for simulating 
the evolution of open quantum systems, the probability for a 
spontaneous 
photo-emission in time $\Delta t$ is~\cite{MCWF,Carm}
\begin{equation}
  \gamma |a_1(t)|^2 \Delta t ,
\end{equation}
where $\gamma$ is the spontaneous emission rate of level 1. First 
let $P(t)$ be the
probability that there were no emissions in the interval $[0,t]$, 
given that there was
an emission at time $t=0$. The probability that there is no 
emission in the interval
$[0,t+\Delta t]$ is therefore
\begin{equation}
 P(t+\Delta t) =  P(t)(1-\gamma |a_1(t)|^2 \Delta t) .
\end{equation}
Taking the limit as $\Delta t$ tends to zero $0$ we arrive at a 
differential equation
for P(t), given by
\begin{equation}
 \frac{dP(t)}{dt} = -\gamma |a_1(t)|^2P(t) , \label{defp}
\end{equation}
the solution of which is
\begin{equation}
 P(t) = e^{-\gamma \int_0^t |a_1(t')|^2\; dt'} . \label{eqp}
\end{equation}
The waiting-time distribution, $w(t)$, is then $P(t)$ multiplied by 
the rate of emission at time $t$. It is clear from Eq.(\ref{defp}) 
that this may be
written
\begin{equation}
  w(t) = -\frac{dP(t)}{dt} = \gamma |a_1(t')|^2 e^{-\gamma \int_0^t |a_1(t')|^2\; dt'} . \label{eqw}
\end{equation}
Clearly this equation will provide an analytic solution
for $w(t)$ as 
long as the integral of $a_1(t)$ possesses an analytic solution. 
Naturally this is true
for any truncated Fourier series. In Fig.(\ref{fig3}) we plot the 
waiting-time
distributions that would be generated by systems which produce 
an evolution 
matching the first $n$ terms in the Fourier series given by  
Eq.(\ref{fser}), for
various values of $n$. We calculate $w(t)$ by numerically evaluating Eq.(\ref{eqw}), where $a(t)$ is given by Eq.(\ref{fser}).  Clearly the waiting-time distribution is increasingly peaked for 
increasing $n$. Experimentally there is a great deal of freedom in choosing the 
generalised Rabi frequency $\Omega$ since it is determined by the 
intensity of the laser driving fields. For the plots in Fig.(\ref{fig3}) 
we have used $\Omega=\gamma /100$ which is well within experimental limits. 

Another quantity of interest is the second order correlation 
function of the 
fluorescent light, $G^2(\tau)$. This is the probability density for 
one photon to be
emitted at time $t$ and another photon to be emitted at time 
$t+\tau$, regardless of
how many other photons where emitted in the intervening time 
interval $(t,t+\tau)$.
The second order correlation function is given by $rQ(\tau)$, 
where $Q(\tau)$ is the
probability that a photon is emitted at a time $t+\tau$, given that 
we have a photon
emitted at time $t$, and $r$ is the probability per unit time for a 
photo-emission.
The latter is simply the average rate of photo-emission, which is
\begin{equation}
  r =  \left( \int_0^\infty \!\!\!\! w(t)t\; dt\right)^{-1}.
\end{equation}

To calculate $Q$ we may use $w(t)$, but we must sum over the 
probabilities for
different numbers of photons to be emitted in the interval 
$(t,t+\tau)$~\cite{KKW}.
This is performed very neatly by the formula
\begin{equation}
  Q(\tau) = w(\tau) + \int_0^\tau Q(t)w(\tau-t)\; dt . \label{neatf}
\end{equation}
\setlength{\parindent}{0em}
If we define $\tilde{Q}(z)$ and $\tilde{w}(z)$ as the Laplace 
transforms of $Q(\tau)$
and $w(\tau)$ respectively, then this equation may be expressed 
as
\begin{equation}
  \tilde{Q}(z) = \frac{\tilde{w}(z)}{1-\tilde{w}(z)} .
\end{equation}
For a given $w(t)$ we may therefore obtain the corresponding 
$Q(t)$, and therefore 
$G^2(t)$, by solving Eq.(\ref{neatf}) using Laplace transforms. 
Alternatively the
second order correlation function may be obtained by solving the 
master equation
describing the atomic system, and using the quantum regression
theorem~\cite{Carm}. For our purposes it is simplest to obtain 
$Q(\tau)$, and hence
$G^2(\tau)$, by numerically integrating Eq.(\ref{neatf}).
\setlength{\parindent}{1em}

\begin{figure}
\begin{center}
\leavevmode
\epsfxsize=6cm 
\epsfbox{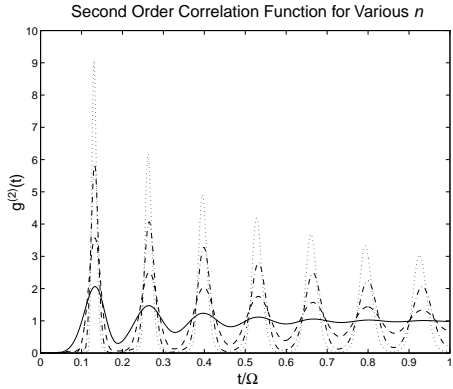}
\caption{\narrowtext The normalised second order correlation function that 
would be generated by 
systems which produce an evolution matching the first $n$ terms 
in the Fourier series given by Eq.(\protect\ref{fser}), for various 
values of 
$n$. Solid line: $n=2$. Dashed line: $n=4$. Dash-dot line: $n=8$. 
Dotted line: $n=16$. The value of the spontaneous emission rate is
$\gamma=100\Omega$. }
\label{fig4}
\end{center}
\end{figure}

In Fig.(\ref{fig4}) we plot the normalised second order correlation 
function, 
$g^2(\tau) = G^2(\tau)/r^2$, for various values of $n$. As we 
expect, this has a series
of peaks and troughs which gradually even out with time. As the 
regularity of the
photon emission increases, longer separation times, $\tau$, are 
required before the
correlation function becomes smoothed out.

We have considered in detail here only the ideal case in which spontaneous emission from atomic states other than level one can be ignored. However, the effects of spontaneous emission from the other atomic levels is easily estimated. In this scheme the evolution of the amplitude of level one, $a_1(t)$, and the ratio $\gamma/\Omega$ are chosen so that, excluding spontaneous emission from any other atomic level, level one will emit after a time delay of approximately $1/(8\Omega)$ (that is, approximately one eighth through the generalised Rabi cycle which has period $1/\Omega$). Clearly we may safely over estimate the amplitude of the other levels by taking them to be unity. The effect of another state with a decay rate of $\gamma_o=(8\Omega)/k$, where $k$ is some dimensionless number, is therefore to
\setlength{\parindent}{0em}

interrupt the normal (regular) spontaneous emission sequence on average no more than once every $k$ emissions. Taking $\gamma = 100\Omega$, which is the value we have used for the previous plots, then for a value of $\gamma_o=\gamma/10^{4}$, spontaneous emissions from this level will, on average, interrupt the desired sequence fewer than approximately once every 1000 emissions. Clearly in this case, and for larger ratios $k$, the waiting time distributions we have calculated will be a good approximation to the true distributions; any changes to the curves will be on the order of $1/1000$ or less of the height of the main peak.\setlength{\parindent}{1em}

Clearly to implement our scheme experimentaly requires that, for a given atom, atomic levels can be found with very different Spontaneous emission rates. Hovever, this is not dificult, as spontaneous emmision rates for atomic levels are found to vary by many orders of magnitude. Observation of atomic shelving~\cite{CKetc} (interruptions of the fluorescence from one `bright' level by shelving to a meta-stable level) with single trapped ions has already been performed~\cite{BNS}. In these experiments the spontaneous emission rate of the meta-stable level is of the order of $10^{-8}$ of that of the bright level. An experimental realisation of our scheme, at least for the four level system which we have treated explicitly, therefore appears to be well within the limits of current technology.

\section{Conclusion} \label{sec5}
We have shown that multiply coherently driven multi-level 
atomic systems 
may be designed to produce light which is more anti-bunched 
than for a two level system. We have given a procedure for 
deriving the driving required to produce highly anti-bunched 
light for a given number of atomic levels. This procedure requires 
that one  of the
atomic levels have a spontaneous emission rate which is much 
larger than the
spontaneous emission rates of the other levels. We have also 
shown that as the
number of levels is increased, the fluorescent light may be made 
increasingly
regular.

We have treated explicitly the case of a four-level system, 
calculating a possible driving configuration to produce highly anti-bunched light.

\section*{Acknowledgements}
We would like to thank H. M. Wiseman and P. L. Knight for helpful 
comments.
We would like to acknowledge support from the University of 
Auckland, the
Association of Commonwealth Universities and the New Zealand 
Vice-Chancellor's
Committee.

\end{multicols}
\widetext
\begin{multicols}{2}

\end{multicols}

\end{document}